\documentclass[twocolumn,twoside,slac_two]{revtex4}
\usepackage{graphicx}
\usepackage{fancyhdr}
\usepackage{graphics}
\usepackage{epstopdf}
\usepackage{textpos}
\begin{document}

\title{\centering 
Measurement of tau parameters and mu-tau universality tests}


\author{
\centering
\includegraphics[scale=0.15]{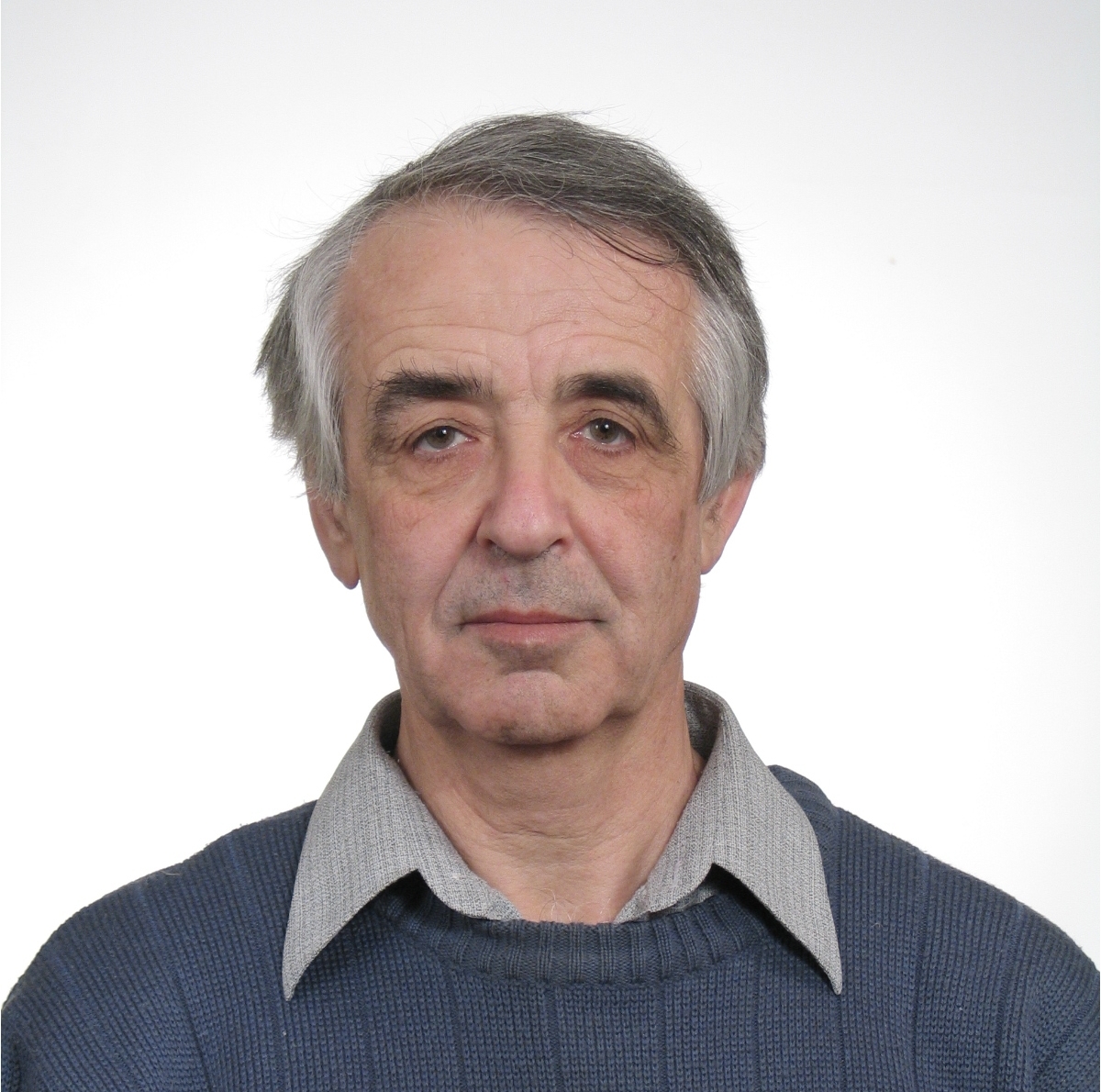} \\
\begin{center}
B.A.Shwartz
\end{center}}
\affiliation{\centering 
Budker Institute of Nuclear Physics of SB RAS, Novosibirsk, 630090, Russia }
\begin{abstract}
The report reviews the measurements of the tau lepton parameters. The tau mass
measurements at the KEDR detector as well as at the B-factories are considered
in more details. The present limitations on the lepton universality tests are
discussed. 
\end{abstract}

\maketitle
\thispagestyle{fancy}

\section{Introduction}

According to the Standard Model (SM) the cupling of W-boson with leptons is family-independent:  $g_e = g_{\mu} = g_{\tau}$ (lepton universality).
Thus, all lepton decays induced by the charged currents are governed by the same
weak constant:
\[ G_F = \frac{g^2}{4\sqrt(2)M_W^2}, \quad g= g_e = g_{\mu} = g_{\tau},   \]
where $G_F$ is Fermi constant.

According to the present knowledge $g_e = g_{\mu}$ within at least 0.2\% while
an upper limit on the difference  $(g_{\mu} - g_{\tau})$ is 2\% \cite{pich08}.
The natural idea is to check the lepton universality using ratios of the tau 
branching fractions.
Tests of the $g_l$ equality were performed in the $W$ decays 
(ALEPH, DELPHY, L3 and OPAL),
$\tau$ decays (ALEPH, DELPHY, L3, OPAL and CLEO),
kaon decays (KLOE) and pion decays (TRIUMPH and PSI).

This report describes the status of the measurements of the tau parameters 
relevant to the lepton universality check and the status of these tests. 

\section{Tau mass measurement at the VEPP-4M collider with the
KEDR detector}

To test the $g_{\tau}/g_{\mu}$ ratio the precise value of $\tau$ mass
is important. Recently, new measurements of this parameter were made
by the KEDR, Belle and BaBar experiments.

In the KEDR experiment \cite{kedr-det}, 
which operates at the VEPP-4M collider at BINP, the $\tau$-lepton mass
was derived from the measurements of $e^+e^- \to \tau^+\tau^-$ cross section
near threshold \cite{kedr-tau-mass}.
The key problem for this approach is the precise energy determination.
Two independent methods were used in this experiment.
One of them was resonant depolarization method which provided the accuracy of
the energy determination of about 10~keV. The other one used Compton 
backscattering of the laser photons by the beam in the collider and determined
the energy with the accuracy 50-70~keV.

\begin{figure}[hbtp]
\centering
\includegraphics[width=0.5\textwidth]{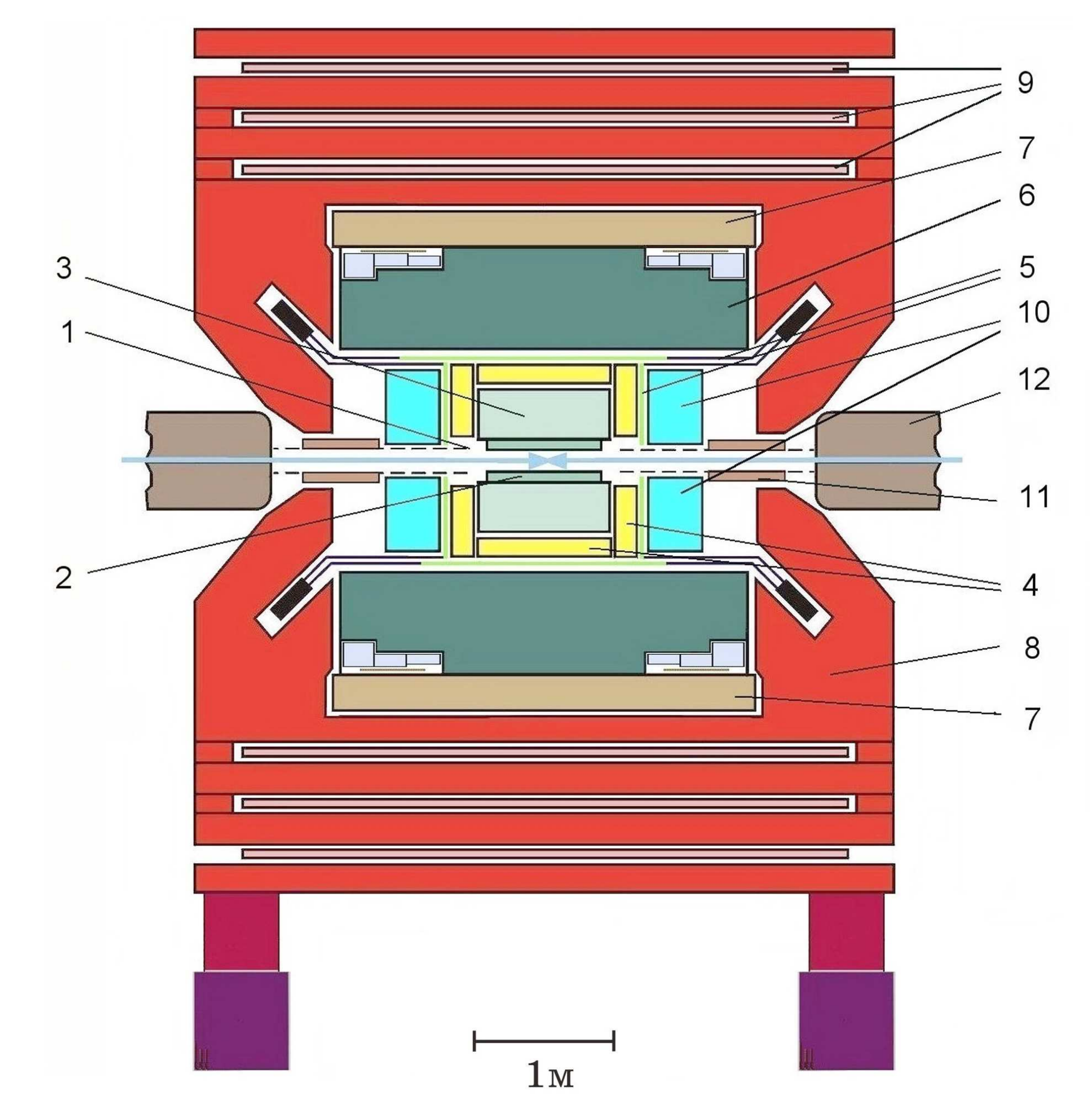}
\caption{The KEDR detector. 1 -- beam pipe;
2 -- vertex detector;
3 -- drift chamber;
4 -- aerogel Cherenkov counters;
5 -- time-of-flight scintillation counters;
6 -- liquid krypton calorimeter;
7 -- superconducting solenoid;
8 -- iron yoke;
9 -- muon system;
10 -- CsI endcap calorimeter;
11 -- compensating solenoids;
12 -- quadrupole lenses.}
\label{fig:kedr-scheme}
\end{figure}

A schematic view of the KEDR detector is shown in Fig.~\ref{fig:kedr-scheme}.
The KEDR detector comprises a vertex detector, a drift chamber, 
a time-of-flight system of scintillation counters, a particle 
identification system based on aerogel Cherenkov counters, a calorimeter 
with the longitudinal segmentation (liquid krypton in 
the barrel part and CsI crystals in the end caps) and 
a muon tube system inside the magnet yoke. 
An axial magnetic field of 0.6 Tesla is produced by the
superconducting solenoid.

To diminish systematic uncertainties the event selection
 criteria were chosen as loose as possible while a background
 was kept to be negligible. The two-prong events with up to
three photons due to 
$e^+e^- \to (\tau \to e\nu\nu)( \tau \to e\nu\nu , µ\nu\nu, 
\pi\nu, K\nu, \rho\nu)^* $ + c.c. 
were selected. At least one track must be identified as 
an electron using the signal in the calorimeter and the 
momentum measurement. The $\mu/\pi/K$ identification 
was not applied as it does not reduce the systematic uncertainty 
of the mass. No photons with $E_{\gamma}>$30~MeV were allowed.
The other cuts were $E<$2200~MeV, $p_T>$200~MeV, $p_T/(W-E)>0.06$,  
where $p_T$ is the total transverse momentum, $E$ is total energy
of the detected particles and $W = 2E$ beam . With these cuts
the residual background (mainly two-photon) is expected to be 
uniform in the energy region of the experiment. 

\begin{figure}[hbtp]
\centering
\includegraphics[width=0.5\textwidth]{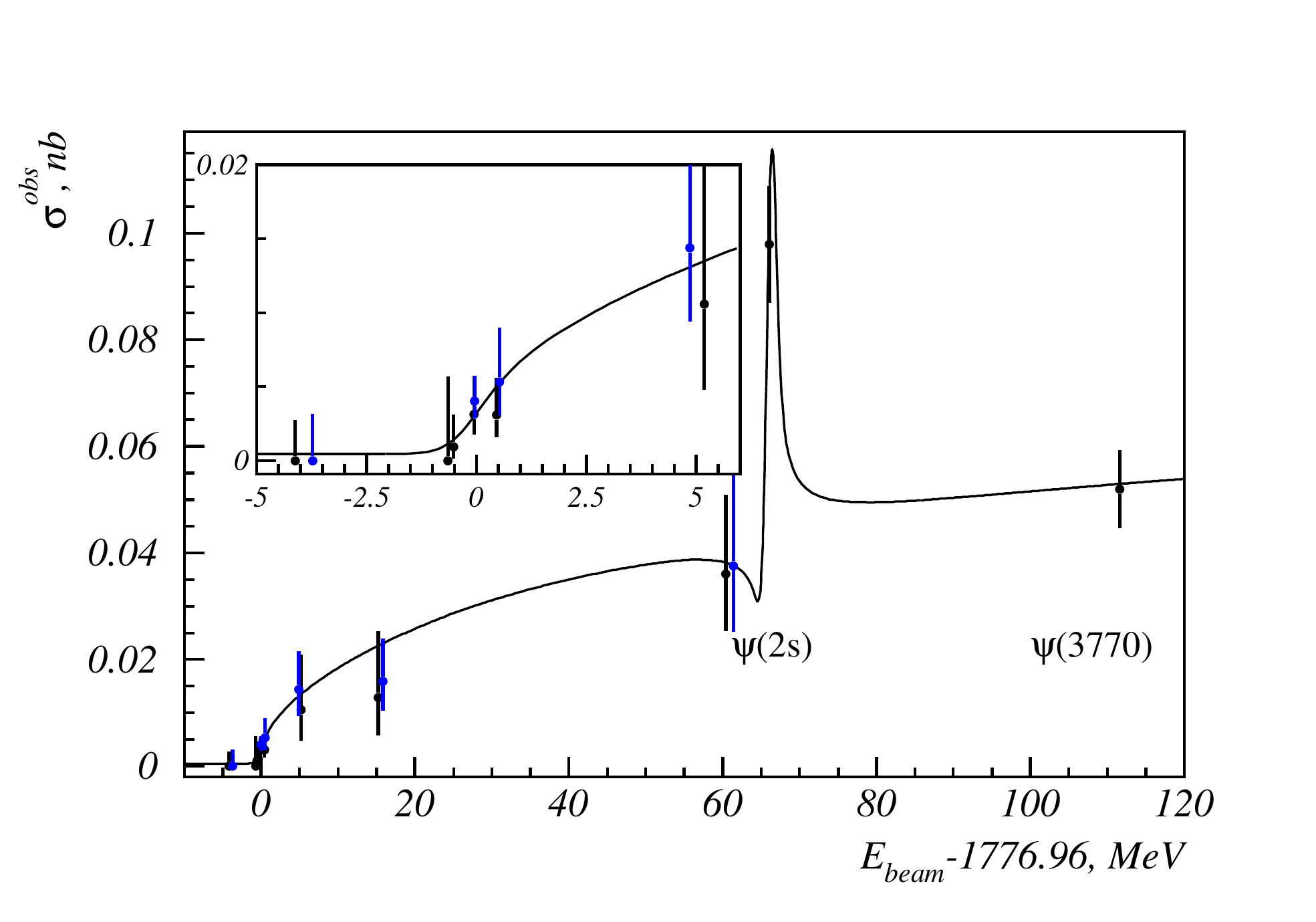}
\caption{$\tau^+\tau^-$ cross section measured by the KEDR detector.}
\label{fig:kedr-tau}
\end{figure}

An integrated luminosity of $L=14.3$~pb$^{-1}$ provided 
26 events at the threshold range. Measured $\tau^+\tau^-$ cross section 
is shown in the Fig.~\ref{fig:kedr-tau}.
A fit of the data produced in the following result:
\[M_{\tau} = 1776^{+0.17}_{-0.19} \pm 0.15 \mathrm{MeV}, \]
where the first error is statistical and the second one is systematic.

Conservative estimations of the systematic uncertainties are
presented in the Table~\ref{tab:sys-err}.
\begin{table}[htbp]
	\centering
\caption{Systematic uncertainties.}
		\begin{tabular}{lc}
\hline 
Beam energy determination & 				35 keV \\
Detection efficiency variations & 			120 keV \\
Energy spread determination accuracy & 		20 keV \\
Background dependence on the beam energy & 	20 keV \\
Luminosity measurement instability & 		80 keV  \\
Beam energy spread variation & 			10 keV  \\
Cross section calculation (r.c.,  0 interference) & 	30 keV \\ \hline
Sum in quadrature & 				150 keV		\\	\hline
		\end{tabular}
	\label{tab:sys-err}
\end{table}

\section{Tau mass measurement at Belle and BaBar }

During last ten years a lot of physics results came from two B-factories –- 
Belle~\cite{belle}   and BaBar~\cite{babar}.
Both detectors are forward/backward asymmetric detectors
with high vertex resolution, magnetic spectrometry, excellent calorimetry  
and sophisticated particle identification ability. 
Integrated luminosity collected by both detectors exceeded 1500~fb$^{-1}$.

\begin{figure}[hbtp]
\centering
\includegraphics[width=0.5\textwidth]{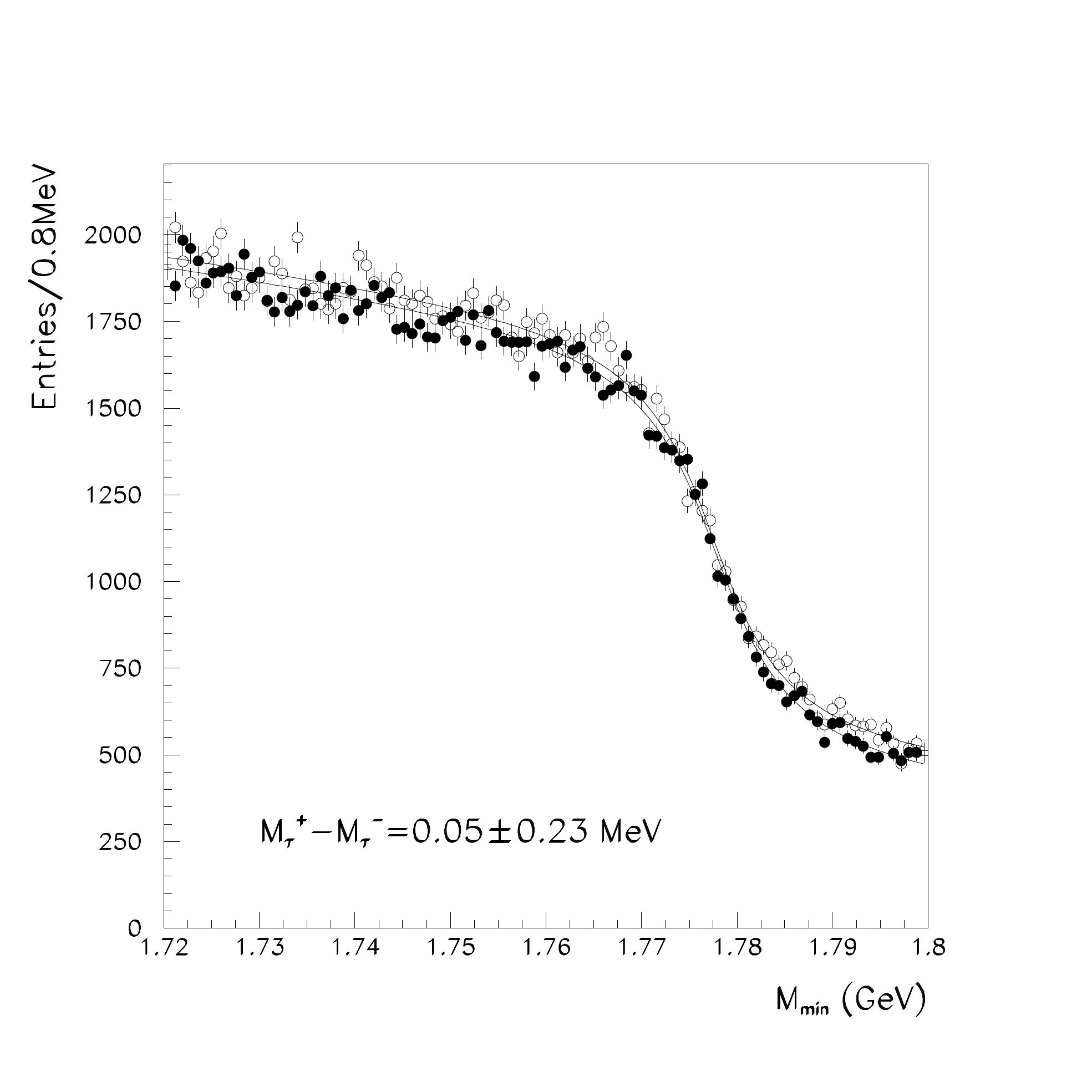}
\caption{Pseudomass distribution, Belle experiment. The black and empty
circles correspond to the negative and positive tau leptons. }
\label{fig:bel-pmass}
\end{figure}

In the Belle and BaBar experiments the $\tau$ mass values were determined by the
fit of the pseudomass distribution of the hadronic decays. The pseudomass 
is defined by the formulae: \\
\begin{math}
M_{\tau}^2  =  (E_h+E_{\nu})^2-(\overrightarrow{p}_h + 
\overrightarrow{p}_{\nu})^2 = \\
  M_h^2 + 2(E_{\tau}-E_h)(E_h-p_h\cos(\theta)) \geq 
M_p^2 = \\ M_h^2 + 2(E_{\tau}-E_h)(E_h-p_h).  	
\end{math}
\\
Typical pseudomass spectrum obtained in~\cite{belle} is shown in the 
Fig.~\ref{fig:bel-pmass}.
The results of the BaBar and Belle experiments, based on 
389 and 370 million $\tau^+\tau^-$ pairs respectively 
\cite{babar-mt,belle-mt} are presented in the Table~\ref{tab:tau-mass}.

\begin{table}[hbtp]
	\centering
	\caption{Belle and BaBar tau mass measurements}
		\begin{tabular}{lcc}
\hline 
Group & BaBar & Belle \\ \hline 		
 $\int Ldt$, fb$^{-1}$ & 423 & 414 \\ \hline 
$N_{\tau\tau}, 10^6$	& 388	& 380 \\ \hline
$N_{ev}, 10^5$ &	682 &	580 \\ \hline
$M_{\tau}$, MeV &1776.68$\pm$0.12$\pm$0.41 &	1776.61$\pm$0.13$\pm$0.35 \\ \hline
		\end{tabular}
	\label{tab:tau-mass}
\end{table}

\begin{figure}[hbtp]
\centering
\includegraphics[width=0.5\textwidth]{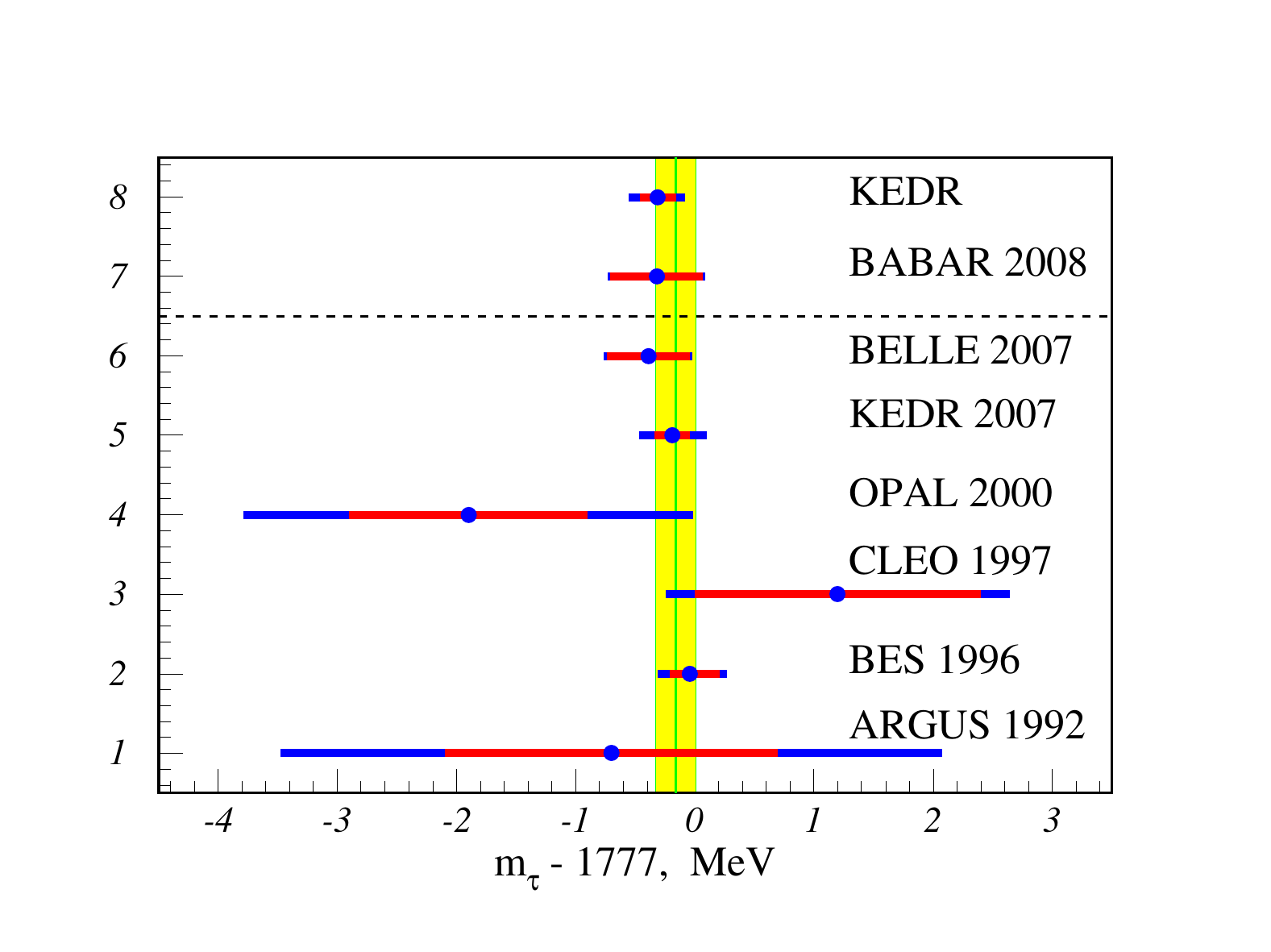}
\caption{Measurements of $\tau$-lepton mass.}
\label{fig:all-tau}
\end{figure}

The progress of $M_{\tau}$ measurements can be seen in the
Fig.~\ref{fig:all-tau}. 
The quoted results should be compared with the PDG average \cite{pdg2010}:
$M_{\tau}=(1776.82 \pm 0.16)$~MeV. 

The pseudomass method provides a separate results for the masses of the positive
and negative leptons. The difference between these masses  tests CPT theorem.
The results are presented in the Table~\ref{tab:m-dif}.
\begin{table}[hbt]
\centering
\caption{Experimental values of the $\Delta m = m_{\tau^+}-m_{\tau^-}$}
\begin{tabular}{|c|c|c|c|}
\hline
Experiment & OPAL,  & Belle,  & BaBar, \\ 
  & 2000 & 2007 & 2008  \\ \hline
$N_{\tau^+\tau^-}, 10^6$       &  0.16  & 380  &  388  \\
$ \Delta m/m_{\tau}, 10^{-4} $ &  $0.0\pm 18.0$ & $0.3\pm 1.5$ & $-3.4\pm 1.3$ 
 \\
$ \Delta m/m_{\tau}, 10^{-4} $, 90\% CL & $<30.0$ & $< 2.8$ & $ < 5.6 $ \\ \hline   
\end{tabular}
\label{tab:m-dif}
\end{table}

After all efforts on tau mass measurement, its contribution to the lepton 
universality check became very small and now the accuracies of other tau 
parameters dominate. The ratio $r = g_{\tau}/g_{\mu}$ is expressed as:

\begin{equation}
r= \left( \frac{g_{\tau}}{g_{\mu}} \right)^2 =
\left( \frac{m_{\mu}}{m_{\tau}} \right)^5
\left( \frac{\tau_{\mu}}{\tau_{\tau}} \right)
\frac{F_{cor}(m_{\mu},m_e)}{F_{cor}(m_{\tau},m_e)}, 	
\end{equation}
where a function $F_{cor}(m_1,m_2)$ takes into account the electroweak 
corrections. These corrections are small and are calculated with high
accuracy. Contributions of the uncertainties of the relevant parameters
to the final precision of this ratio can be seen in the Table~\ref{tab:r}. 

\begin{table}[hbt]
\centering
\caption{Measurement of the ratio $r=g_{\tau}/g_{\mu}$}
\begin{tabular}{|c|c|c|c|}
\hline
$r$ & $tau_{\tau}$, fs	 & $B(\tau \nu_e\nu_{\tau})$,\% & $m_{\tau}$ ,MeV \\ 
\hline
1.0034	& $ 290.6 \pm 1.0$ & $17.85 \pm 0.05$ & $1776.82 \pm 0.16$ \\ \hline
  $\pm 0.0045$  &  $\pm 0.0034$  & $\pm 0.0028$ &  	$\pm 0.0004$ \\ \hline
\end{tabular}

\label{tab:r}
\end{table}

\section{Measurements of  tau lifetime}

The most precise published results on the tau lepton lifetime were obtained by
the CLEO (1996), OPAL (1996), ALEPH (1997), L3 (2000) and DELPHY (2004).
The typical total (statistical plus systematics) precisions were (2-4)~fs.
However, averaged by PDG value has much better accuracy:   
\cite{pdg2010} $\tau_{\tau}$=290.6$\pm$1.0~fs.  

The Belle and the BaBar collected a huge statistics of the tau lepton decays.
However, the tau lepton energy in these experiments is much lower than that was
at LEP, which implies much shorter decay length and, as a result, much
more complicated analysis. The BaBar collaboration released a preliminary
result several years ago \cite{babar-tautau},
$\tau_{\tau}$ = 289.40$\pm$0.91(stat.)$\pm$0.90(syst.)~fs,
but final results are not published yet.

The tau lifetime analysis at the Belle is in progress \cite{belle-tautau}. 
The result, even preliminary, is not released yet, 
however, the achieved accuracy referred as ($\pm 0.37$(stat)$\pm 0.33$(syst))~fs. 
Analysis of systematics is in progress.

Since the lifetime of the positive and negative tau leptons
are measured separately, the difference between them can be derived. 
The Belle experiment measured this difference using the decays 
 $e^+e^- \to \tau^+\tau^- \to (3\pi\nu)(3\pi\nu)$:
$\Delta(c\tau ) = c\tau^+ - c\tau^- = 0.16 \pm 0.22\mu$m that corresponds to
$|\tau^+ - \tau^- |/\tau_{av} < 6 \times 10^{-3}$  @ 90\% CL.

The BaBar result on this difference is \cite{babar-tautau}:
$ (\tau^+ - \tau^-)/(\tau^+ + \tau^-) =(0.12 \pm 0.32)$\%,

\section{Lepton Universality and Branching Fractions}

Lepton branching fractions of the tau lepton are very important for the lepton
universality check. The values provided by PDG \cite{pdg2010},
$B_{\mu\nu\nu} = 17.36 \pm 0.05 $\% and $B_{e\nu\nu} = 17.85 \pm 0.05$\%,
already have high accuracy and improvement of the precision is a quite 
difficult task.
However, the ratio of the branching fractions can be measured with better 
accuracy using the large statistics collected at the B-factories. Recently the
BaBar collaboration measured several ratios~\cite{babar-br} 
used 467~$fb^{-1}$ of the integrated luminosity:
$\frac{B(\tau^- \to \mu^- \bar{\nu}_{\mu} \nu_{\tau})}{B(\tau^- \to e^- \bar{\nu}_e \nu_{\tau})} = (0.9796 \pm 0.0016 \pm 0.0036)$,
$\frac{B(\tau^- \to \pi^- \nu_{\tau})}{B(\tau^- \to e^- \bar{\nu}_e \nu_{\tau})} = (0.5945 \pm 0.0014 \pm 0.0061)$, and
 $\frac{B(\tau^- \to K^-  \nu_{\tau})}{B(\tau^- \to e^- \bar{\nu}_e \nu_{\tau})} = (0.03882 \pm 0.00032 \pm 0.00057)$,
 where the uncertainties are statistical and systematic, respectively.

The first measured ratio provides a check of muon/electron universality:
$(g_{\mu}/g_e) = 1.0036 \pm 0.0020$ which does not contradict to SM and
improves slightly the averaged PDG value.
Two latter ratios can be used for $(g_{\tau}/g_{\mu})$ check taking
into account branching fractions $\pi \to \mu \nu_{\mu} $ and
$K \to \mu\nu_{\mu} $. The BaBar presented the following combined result:
$(g_{\tau}/g_{\mu}) = 0.9850 \pm 0.0057$ that is 2.8 standard deviation 
lower than SM prediction.

\section{Project of tau mass measurement at BEPC II}

Few years ago the modified collider, BEPC-II with the new detector, BES-III,
came to the operation at the IHEP in Beijing~\cite{bes3}. 
Main parameters of the collider are presented in the Table~
\begin{table}[hbtp]
\centering
\caption{Main parameters of the BEPC collider}
\begin{tabular}{lc}
\hline
Beam energy    	&   1.0-2.3~GeV   \\
Luminosity      	&   $10^{33}$~cm$^{-2}$s$^{-1}$  \\
Optimum energy		&     1.89~GeV  \\
Energy spread		&   $5.16\times 10^{-4}$  \\
No. of bunches		&         93  \\
Bunch length		&			1.5~cm  \\ 
Total current		&      0.91~A   \\ \hline
\end{tabular}
\label{tab:bepc}
\end{table}

\begin{figure}[hbtp]
\centering
\includegraphics[width=0.5\textwidth]{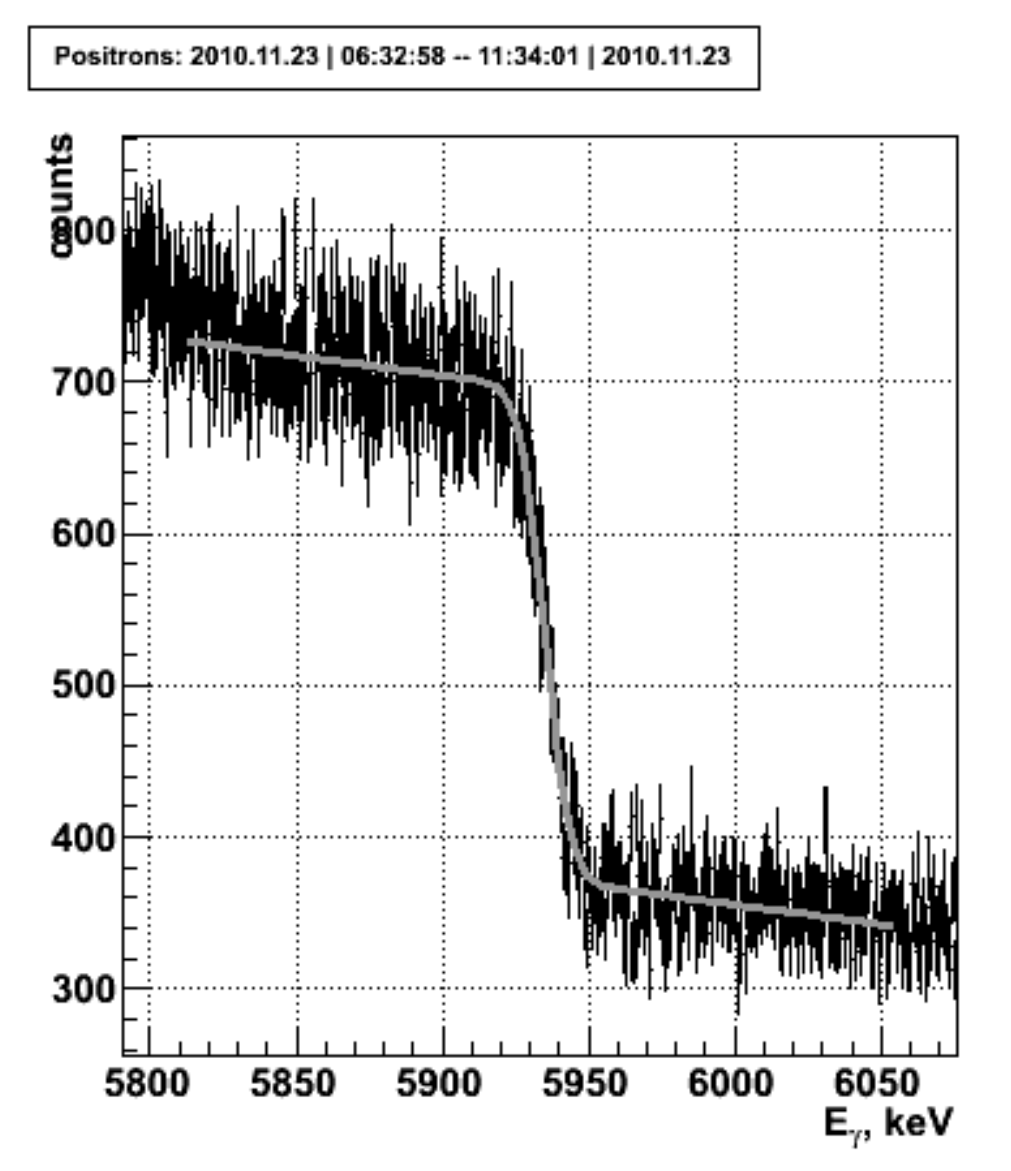}
\caption{Backscattered photon energy spectrum \cite{bes-en} . The solid line 
is a fit.}
\label{fig:compton}
\end{figure}

One of the important studies planning at BES-III is tau lepton mass 
measurement with the accuracy better than 100~keV~\cite{bes-en}  .
The beam energy measurement system based on the Compton backscattering of the
laser photons by the beam has been created for this experiment. 
First test measurements were performed in the J/$\Psi$ energy range.
The backscattered photon energy distribution at the end of the Compton
spectrum is shown in the Fig.~\ref{fig:compton}.
The position of the Compton edge can be precisely determined. 

\begin{figure}[hbtp]
\centering
\includegraphics[width=0.5\textwidth]{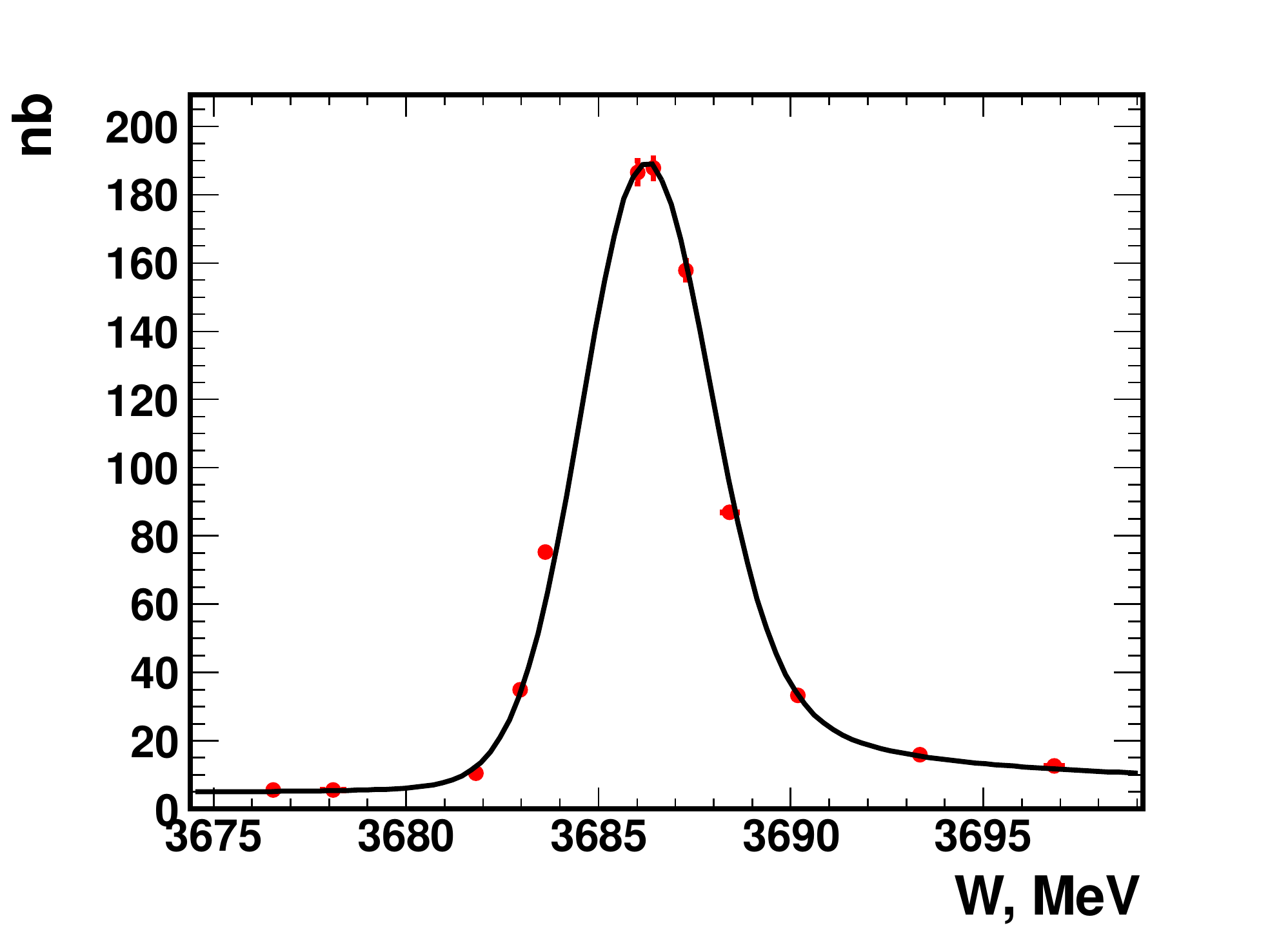}
\caption{Cross section test measurement at $J/\Psi$ - meson \cite{bes-en}.}
\label{fig:test-psi}
\end{figure}

The J/$\Psi$ curve measured using the energy measurement system is 
presented in Fig.~\ref{fig:test-psi}.
The obtained systematic uncertainty of the beam energy is 
$\delta E/E \approx 2 \times 10^{-5}$.

\section{Results from LHC experiments}

Recently started experiments at LHC have already collected large number
of events with $W$ bosons production.
In general, a measurement of the leptonic decays of the $W$ bosons provides
a direct check of the lepton universality.   
This year the ATLAS collaboration presented the measurement of the 
$W$ boson production cross section followed with the leptonic decay 
\cite{atlas-w1,atlas-w}.
Obtained cross section value with the decay 
$W \to \tau \nu_{\tau}$ is: 
\[\sigma^{tot}_{W \to \tau\nu} = (11.1 \pm 0.3 \mathrm{(stat)} 
\pm 1.7 \mathrm{(syst)} \pm 0.4 \mathrm{(lumi))\, nb},
\]
where the first error is statistical, second is systematic and the third one
is induced by the uncertainty of the luminosity determination.
The CMS collaboration also observed such decays \cite{cms-w} but the values
are not released yet.    

\begin{figure}[hbtp]
\centering
\includegraphics[width=0.5\textwidth]{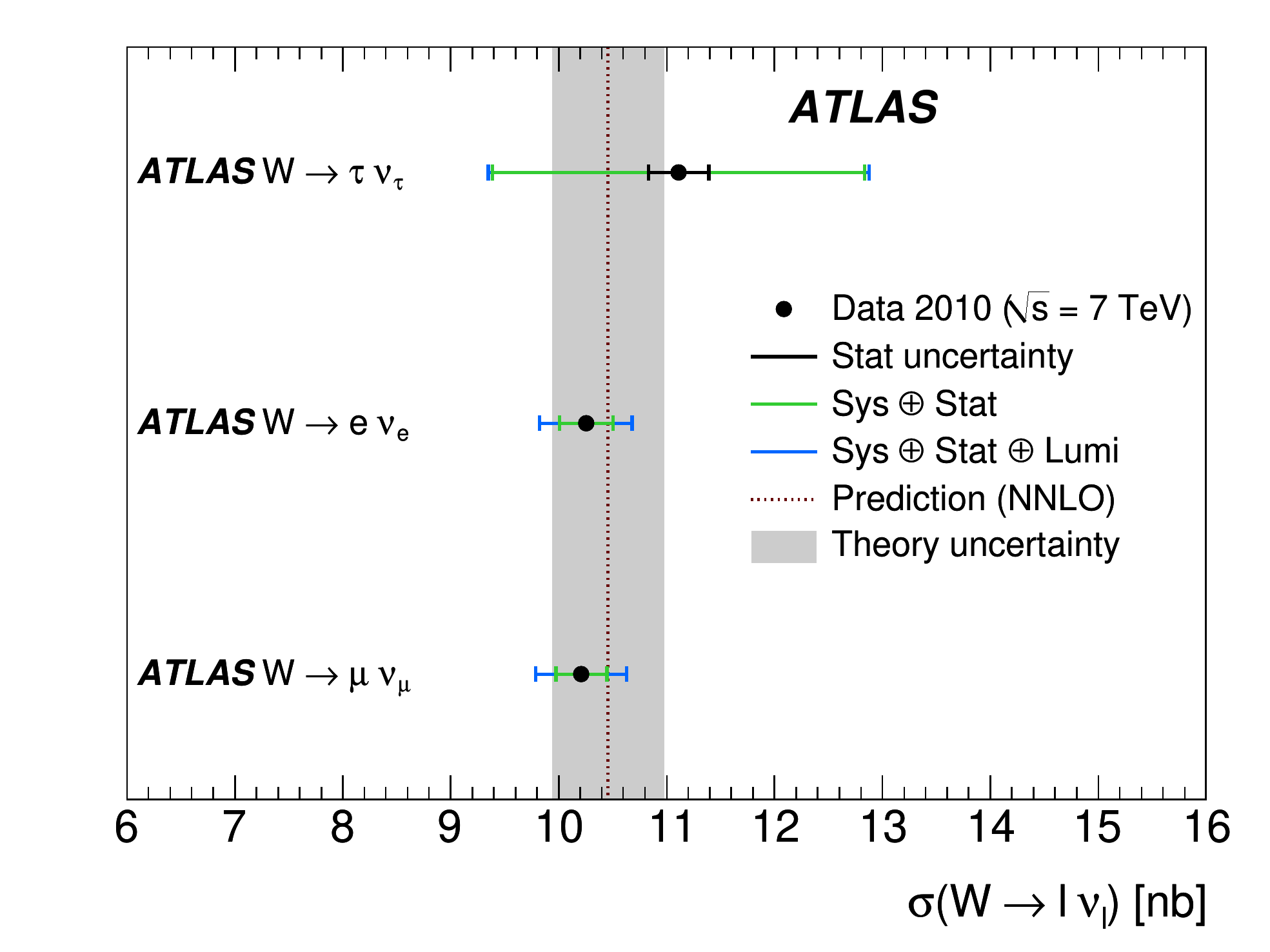}
\caption{Cross sections for the different W leptonic channels measured
by the ATLAS \cite{atlas-w}.}
\label{fig:atlas-lept}
\end{figure}

The results from ATLAS are presented in Fig.~\ref{fig:atlas-lept} taken from
\cite{atlas-w}. 
As seen from the figure, accuracy of the measurements does not allow yet
to improve the check of the lepton universality. However, we can hope on the
improvement of the precision in future.

\section{Conclusion}

\begin{itemize}
    \item In the last decade a lot of efforts were applied to measure 
tau mass with high accuracy.
 At present the most crucial parameters for the lepton 
universality check are branching fractions and tau lifetime. 
    \item To improve accuracy in the $(g_{\tau}/g_e)$ and $(g_{\tau}/g_{\mu})$
ratios, first of all, more precise measurements of the tau leptonic branching
fractions are needed.
Huge data samples of the tau decays ($10^9$ events) is available at 
B-factories. However, to make use from high statistics a decrease of the systematic uncertainties is necessary, which is a very difficult task.
    \item A considerable progress in the tau mass measurement is anticipated
in near future from a new experiment prepared at the BEPC-II/BES-III. 
We can hope for the further progress in the measurements of the tau lepton
parameters in the LHC experiments as well as future super B- and C-factories.

\end{itemize}



\begin{thebibliography}{99}   
%
%
\bibitem{pich08} A.~Pich, Nuclear Physics B (Proc. Suppl.), {\bf 181–182}, 
300 (2008)
%
\bibitem{kedr-det} V.V.~Anashin {\it et~al.}, Nucl. Instr. Meth. A  {\bf 478},
420 (2002)
%
\bibitem{kedr-tau-mass} A.G.~Shamov {\it et~al.}, Nucl. Phys. B (Proc. Suppl.),
{\bf 189}, 21 (2009)
%
\bibitem{belle} A.~Abashian {\it et~al.}, 
Nucl. Instr. Meth. A, {\bf 479}, 117 (2002)
%
\bibitem{babar} B.~Aubert {\it et~al.},
Nucl. Instr. Meth. A, {\bf 479}, 1 (2002)
%
\bibitem{babar-mt}
B.~Aubert {\it et~al.}, Phys. Rev. D, {\bf 80}, 092005 (2009)
%
\bibitem{belle-mt}
K.~Belous {\it et~al.}, Phys. Rev. Lett., {\bf 99}, 011801 (2007)
%
\bibitem{pdg2010} K.~Nakamura {\it et~al.} (Particle Data Group), 
Journal of Physics G, {\bf 37}, 075021 (2010) 
%
\bibitem{babar-tautau} A.~Lusiani, Nuclear Physics B (Proc. Suppl.), 
{\bf 144}, 105 (2005)
%
\bibitem{belle-tautau} S.I.~Eidelman (for the Belle Collaboration),
Nuclear Physics B (Proc. Suppl.), {\bf 218}, 172 (2011)
%
\bibitem{babar-br} B.~Aubert {\it et~al.}, Phys. Rev. Lett., {\bf 105}, 
051602 (2010)
%
\bibitem{bes3} A.~Zhemchugov, 
Nucl. Phys. B (Proc. Suppl.), {\bf 189}, 353 (2009) 
\\
S.L.~Olsen, AIP Conf. Proc., {\bf 1182}, 402 (2009) 
%
\bibitem{bes-en} E.V.~Abakumova {\it et~al.},  
Nucl. Instr. Meth. A, {\bf 659}, 21 (2011)
%
\bibitem{atlas-w1} ATLAS Collaboration, 
ATLAS Conference Note ATLAS-CONF-2011-041 (2011)
%
\bibitem{atlas-w} ATLAS collaboration,  CERN-PH-EP-2011-122, CERN (2011) \\
arXiv:1108.4101v2.
%
\bibitem{cms-w} CMS collaboration, CMS-PAS-EWK-11-002, CERN (2011) 

\end{thebibliography}
\end{document}